\documentstyle[prd,aps,floats]{revtex} 
\draft

\begin{document}

\twocolumn[\hsize\textwidth\columnwidth\hsize\csname
@twocolumnfalse\endcsname

\title{%
\hbox to\hsize{\normalsize\rm December 1996, Final Version March 
1997 \hfil
Preprint TECHNION-PH-96-23}
\hbox to\hsize{\normalsize\rm To be published in Physical 
Review D\hfil  MPI-PhT/96-126}
%\hbox to\hsize{\hfil}
\vskip6pt
Cherenkov radiation by massless neutrinos in a magnetic field}

\author{Ara N.~Ioannisian} 
\address{Yerevan Physics Institute, Alikhanyan Br.~2, Yerevan, 
375036, Armenia}
\address{and Department of Physics, Technion---Israel Institute of 
Technology, Haifa 32000, Israel}

\author{Georg G.~Raffelt}
\address{Max-Planck-Institut f\"ur Physik 
(Werner-Heisenberg-Institut), 
F\"ohringer Ring 6, 80805 M\"unchen, Germany}

\date{8 December 1996}
\maketitle

\begin{abstract}
We calculate the Cherenkov process $\nu\to\nu\gamma$ in the presence
of a homogeneous magnetic field. The neutrinos are taken to be
massless with only standard-model couplings.  The magnetic field
fulfills the dual purpose of inducing an effective neutrino-photon
vertex and of modifying the photon dispersion relation such that the
Cherenkov condition $\omega<|{\bf k}|$ is fulfilled.  Our effect is
closely related to photon splitting that occurs in magnetic fields and
that may be astrophysically important in the strong magnetic fields of
pulsars.  It is also closely related to magnetic-field enhanced
radiative decays $\nu\to\nu'\gamma$ that have been extensively
discussed in the recent literature. In the appropriate limits we agree
with these results, but we disagree with earlier explicit calculations
of the Cherenkov process.  For a field strength $B_{\rm
crit}=m_e^2/e=4.41\times10^{13}~{\rm Gauss}$ and for $E=2m_e$ the
Cherenkov rate is about $6\times10^{-11}~{\rm s}^{-1}$ and thus
too small to be of practical importance for pulsar~physics.
\end{abstract}

\pacs{PACS numbers: 13.15.+g, 14.60.Lm, 97.10.Ld, 97.60.Gb}

\vskip2.0pc]

%%%%%%%%%%%%%%%%%%%%%%%%%%%%%%%%%%%%%%%%%%%%%%%%%%%%%%%%%%%%%%%%%%%%%%
%% Section I %%%%%%%%%%%%%%%%%%%%%%%%%%%%%%%%%%%%%%%%%%%%%%%%%%%%%%%%%
%%%%%%%%%%%%%%%%%%%%%%%%%%%%%%%%%%%%%%%%%%%%%%%%%%%%%%%%%%%%%%%%%%%%%%

\section{Introduction}

In many astrophysical environments the absorption, emission, or
scattering of neutrinos occurs in dense media or in the presence of
strong magnetic fields \cite{Raffelt}. Of particular conceptual
interest are those reactions which have no counterpart in vacuum,
notably the plasmon decay $\gamma\to\bar\nu\nu$ and the Cherenkov
process $\nu\to\nu\gamma$. These reactions do not occur in vacuum
because they are kinematically forbidden and because neutrinos do not
couple to photons. In the presence of a medium or $B$-field, neutrinos
acquire an effective coupling to photons by virtue of intermediate
charged particles. Also, neutrinos may have anomalous electromagnetic
interactions, for example induced by a magnetic dipole moment. In
addition, media or external fields modify the dispersion relations of
all particles so that phase space is opened for neutrino-photon
reactions of the type $1\to 2+3$.

The plasma process $\gamma\to\bar\nu\nu$ was first studied by Adams,
Ruderman, and Woo \cite{Adams} and Zaidi \cite{Zaidi} in order to
calculate stellar energy losses into neutrinos. The
\hbox{$\nu$-$\gamma$}-coupling is enabled by the presence of the
electrons of the background medium, and the process is kinematically
allowed because the photons acquire essentially an effective mass. The
plasma process is the dominant source for neutrinos in many types of
stars and thus is of great practical importance in
astrophysics~\cite{Raffelt}. For that reason it also lends itself to
deriving astrophysical limits on anomalous electromagnetic neutrino
couplings which provide an additional contribution to the
$\nu$-$\gamma$-vertex and thus to the emission
rate~\cite{Raffelt,Bernstein,RaffeltII}.

The presence of a magnetic field induces an effective
$\nu$-$\gamma$-coupling which contributes to the
\hbox{$\gamma\to\bar\nu\nu$} reaction. The resulting decay rate was
calculated by Galtsov and Nikitina~\cite{Galtsov},
Skobelev~\cite{Skobelev76}, and DeRaad, Milton, and Hari
Dass~\cite{Raad}, assuming that phase space is opened by a suitable
medium- or field-induced modification of the photon refractive index.

If neutrinos are exactly massless as we will always assume, and if
medium-induced modifications of their dispersion relation can be
neglected, the photon decay $\gamma\to\bar\nu\nu$ is kinematically
possible whenever the photon four momentum $k=(\omega,{\bf k})$ is
time-like, i.e.\ $k^2={\bf k}^2-\omega^2<0$.\footnote{We always use
the metric ${\rm diag}({-}{+}{+}{+})$ in accordance with much of the
literature on magnetic-field effects on electromagnetic couplings.}
Often the dispersion relation is expressed by $|{\bf k}|=n\omega$ in
terms of the refractive index~$n$. In this language the photon decay
is kinematically possible whenever $n<1$. In stellar plasmas this
condition is usually satisfied, leading to the great practical
importance of the plasma decay process for the physics of stars.

Even in a normal plasma there are electromagnetic excitations which
fulfill the opposite condition $n>1$, namely the longitudinal plasmons
or Langmuir waves $\gamma_{\rm L}$ which do not exist in vacuum.
Their dispersion relation $\omega=f({\bf k})$ ``crosses the light
cone'' at a certain momentum ${\bf k}_c$ so that ${\bf
  k}^2-\omega^2<0$ for $|{\bf k}|<|{\bf k}_c|$ and ${\bf
  k}^2-\omega^2>0$ for $|{\bf k}|>|{\bf k}_c|$.  Thus there is phase
space for the Cherenkov process $\nu\to\nu\gamma_{\rm L}$. Based on
the standard-model interactions with the electrons of the medium,
Tsytovich~\cite{Tsytovich} was the first to calculate this sort of
process. It was later rediscovered by Oraevsky, Semikoz, and
Smorodinsky \cite{Oraevsky}, Sawyer~\cite{Sawyer}, D'Olivo, Nieves,
and Pal~\cite{Olivo}, and Hardy and Melrose~\cite{Hardy96}.  It was
claimed that an intense neutrino beam could be particularly effective
at emitting Langmuir waves by virtue of a nonlinear feed-back
mechansim (Bingham et al.~\cite{Bingham}), with important consequences
for supernova physics. Unfortunately, this spectacular claim is
erroneous because it was based on the assumption of a spurious
phase-coherence of the neutrino states---see Hardy and
Melrose~\cite{Hardy97}.

Neutrinos may also couple to the electromagnetic field by virtue of an
anomalous magnetic dipole moment, a hypothesis advanced a long time
ago to solve the solar neutrino problem by magnetically induced spin
oscillations.  With this motivation in mind, Radomski \cite{Radomski}
calculated the magnetic-moment Cherenkov process, but unsurprisingly
found it too small to reduce the solar neutrino flux by any
significant amount.  Later this process was reconsidered by Grimus and
Neufeld \cite{Grimus} and Mohanty and Samal~\cite{Mohanty}. The
latter group considered the interior of a supernova core where even
transverse plasmons (photons) appear to have a space-like dispersion
relation because it is dominated by the magnetic moments of the
nucleons rather than the charges of the electrons. In this environment
the phase space for the Cherenkov process is large, while there is
none for the photon decay.

Independently of the nature of the photon dispersion relation the
process $\nu\to\nu\gamma$ occurs at the interface of two media with
different refractive indices (transition radiation). With the
assumption of a neutrino magnetic dipole moment this process was
recently studied by Sakuda and Kurihara~\cite{Sakuda} and 
Grimus and Neufeld~\cite{Grimus95}.

We presently extend previous studies of the Cherenkov process to
neutrinos propagating in an external magnetic field. Around pulsars,
for example, field strengths around the critical value $B_{\rm
crit}=m_e^2/e=4.41\times10^{13}~{\rm Gauss}$ and perhaps even larger
are thought to occur. The electron density is probably so small that
the photon dispersion relation is dominated by the magnetic field. The
Cherenkov condition is then satisfied for significant ranges of photon
frequencies. In addition, the magnetic field itself causes an
effective $\nu$-$\gamma$-vertex by standard-model neutrino couplings
to virtual electrons and positrons. Therefore, we study the Cherenkov
effect entirely within the particle-physics standard model.

A detailed literature search\footnote{The literature on neutrino
Cherenkov radiation and related processes is extremely scattered.
Many of the papers quoted here have never been referenced in the other
papers on the same topic. Therefore, it is quite possible that we have
overlooked other relevant works.} reveals that even this process has
been calculated earlier by Galtsov and Nikitina~\cite{Galtsov} and 
Skobelev~\cite{Skobelev76}. However, we do not agree with their
results, which also sheds doubt on their treatment of the
$\gamma\to\bar\nu\nu$ process.

Our work is closely related to a recent series of papers by Gvozdev,
Mikheev, and Vasilevskaya \cite{Gvozdev} and to papers by
Skobelev~\cite{Skobelev95} and Kachelriess and Wunner \cite{Wunner}
who studied the neutrino radiative decay $\nu\to\nu'\gamma$ in the
presence of magnetic fields where $\nu$ and $\nu'$ are different
neutrino flavors which are assumed to mix. This process would proceed
even in the absence of external fields or media.  In our case of
massless unmixed neutrinos the initial and final state in
$\nu\to\nu\gamma$ is the same flavor and the process does not take
place in vacuum. The role of the external field at modifying the
$\nu$-$\gamma$-vertex in our study is however similar to
Refs.~\cite{Gvozdev,Skobelev95,Wunner}.  In addition, for us it is
crucial that the magnetic field modifies the photon dispersion
relation.  In their case the process is kinematically allowed anyhow,
and it depends on the neutrino mass difference if neglecting the
exact photon dispersion relation is justified.  In the appropriate
limits we agree with the results of
Refs.~\cite{Gvozdev,Skobelev95,Wunner}.

Our work is also related to the process of photon splitting that may
occur in magnetic fields as discussed, for example, in
Refs.~\cite{Adler,Splitting}.  In photon splitting the magnetic field
also plays the dual role of providing an effective three-photon vertex
which does not exist in vacuum, and of modifying the dispersion
relation of the differently polarized modes such that
$\gamma\to\gamma\gamma$ becomes kinematically allowed for certain
polarizations of the initial and final states. In fact, photon
splitting could be called ``Cherenkov radiation by photons in magnetic
fields.''

We proceed in Sec.~II by deriving a general expression for the
Cherenkov rate, assuming a general $\nu$-$\gamma$-vertex.  In Sec.~III
we derive the standard-model effective vertex in the presence of a
homogeneous magnetic field. In Sec.~IV we calculate the Cherenkov rate
on the basis of the magnetic-field modified photon dispersion
relation.  In Sec.~V we summarize our findings.

%%%%%%%%%%%%%%%%%%%%%%%%%%%%%%%%%%%%%%%%%%%%%%%%%%%%%%%%%%%%%%%%%%%%%%
%% Section II %%%%%%%%%%%%%%%%%%%%%%%%%%%%%%%%%%%%%%%%%%%%%%%%%%%%%%%%
%%%%%%%%%%%%%%%%%%%%%%%%%%%%%%%%%%%%%%%%%%%%%%%%%%%%%%%%%%%%%%%%%%%%%%

\section{Cherenkov Radiation}

Beginning with a general discussion of the Cherenkov process 
$\nu(p)\to\nu(p')\gamma(k)$ we note that in terms of the matrix
element ${\cal M}$ the transition rate is
\begin{equation}\label{gg}
\Gamma = \frac{1}{(2\pi)^2}\frac{1}{2E}
\sum_{\rm pols.}\int\frac{d^3{\bf k}}{2\omega}
\frac{d^3{\bf p}'}{2E'}\,\delta^4(p-p'-k)\,|{\cal M}|^2.
\end{equation}
Here, $p=(E,{\bf p})$, $p'=(E',{\bf p}')$, and $k=(\omega,{\bf k})$
are the four momenta of the incoming neutrino, outgoing neutrino, and
photon, respectively. The sum is over photon polarizations. It appears
outside of the phase-space integrals because in general the photon
refractive index depends on the photon polarization state.

With the identity
$d^3{\bf p}'/2E'=\int d^4 p'\,\Theta(E')\delta(p^{\prime 2})$ we may 
integrate over $\delta^4(p-p'-k)$ and find
\begin{eqnarray}\label{ga}
\Gamma &=& \frac{1}{32\pi^2 E^2}\sum_{\rm pols.}
\int \frac{|{\bf k}|}{\omega}\,d|{\bf k}|\,
d\varphi\,d\cos\theta\nonumber\\
&&\hskip2em
\times\,\delta\left(\frac{2E\omega+{\bf k}^2-\omega^2}{2E|{\bf k}|}
-\cos\theta\right)\,|{\cal M}|^2, 
\end{eqnarray}
where $\theta$ is the angle between the emitted photon and incoming
neutrino. We have assumed that the neutrino dispersion relation is
precisely light-like so that $p^2=0$ and $E=|{\bf p}|$. The
integration over the azimuthal photon directions $\varphi$ is not yet
carried out because the photon dispersion relation need not be
isotropic.

The $\delta$-function constrains the photon emission angle to have the
value
\begin{equation}\label{emissionangle}
\cos \theta = n^{-1}\,
\left[1+(n^2-1)\frac{\omega}{2E}\right],
\end{equation}
where we have used the photon refractive index $n=k/\omega$.  Because
it is not isotropic, this opening angle of the Cherenkov ``cone''
actually depends on the azimuthal direction $\varphi$.

In a magnetic field the photon refractive index is not isotropic, and
it depends on the photon polarization. According to Adler's classic
paper~\cite{Adler} there are two eigenmodes of photon propagation, one
with the polarization vector parallel ($\parallel$) and one
perpendicular ($\perp$) to the plane containing ${\bf k}$ and 
${\bf B}$.\footnote{Our definition of $\parallel$ and $\perp$ is 
opposite 
to Adler's~\cite{Adler} who used the photon's magnetic-field vector
to define the polarization.} Therefore, we write the refractive
index in the form
\begin{equation}
n_{\parallel,\perp}=1+
\frac{\alpha}{4\pi}\eta_{\parallel,\perp}\sin^2\beta, 
\end{equation}
where $\beta$ is the angle between ${\bf k}$ and ${\bf B}$.  The
numerical coefficients $\eta_{\parallel,\perp}$ depend on $B$,
$\omega$, and $\beta$. 
For $B={\cal O}(B_{\rm crit})$ they are of order
unity. Therefore, for all situations of practical interest we have
$|n_{\parallel,\perp}-1|\ll 1$. This allows us to expand 
Eq.~(\ref{emissionangle}) to lowest order in $\alpha$,
\begin{equation}\label{par}
\cos\theta= 
1-\frac{\alpha}{4\pi}\eta_{\parallel,\perp}\,
\left(1-\frac{\omega}{E}\right)\,\sin^2\beta.
\end{equation}
This result reveals that to lowest order the outgoing photon
propagates parallel to the original neutrino direction.

Therefore, to lowest order the azimuthal dependence of Eq.~(\ref{ga})
drops out, allowing us to perform both angular integrations
explicitly. Moreover, to this order we do not need to distinguish
between $\omega$ and $|{\bf k}|=n\omega=
\hbox{$\omega[1+{\cal O}(\alpha)]$}$. 
Therefore, to lowest order in $\alpha$ the Cherenkov rate
Eq.~(\ref{ga}) is found to be
\begin{equation}
\label{gc}
\Gamma = \frac{1}{16\pi E^2}\int_0^{\omega_{\rm max}}d\omega
\sum_{\rm pols.}|{\cal M}|^2.
\end{equation}
Energy conservation requires $\omega<E$ so that 
$\omega_{\rm max}\leq E$. The photon dispersion relation ``crosses the
light cone'' at some frequency $\omega_c$ so that the Cherenkov
condition is only satisfied for $0<\omega<\omega_c$. Therefore,
$\omega_{\rm max}=\min(E,\omega_c)$.

%%%%%%%%%%%%%%%%%%%%%%%%%%%%%%%%%%%%%%%%%%%%%%%%%%%%%%%%%%%%%%%%%%%%%%
%% Section III %%%%%%%%%%%%%%%%%%%%%%%%%%%%%%%%%%%%%%%%%%%%%%%%%%%%%%%
%%%%%%%%%%%%%%%%%%%%%%%%%%%%%%%%%%%%%%%%%%%%%%%%%%%%%%%%%%%%%%%%%%%%%%

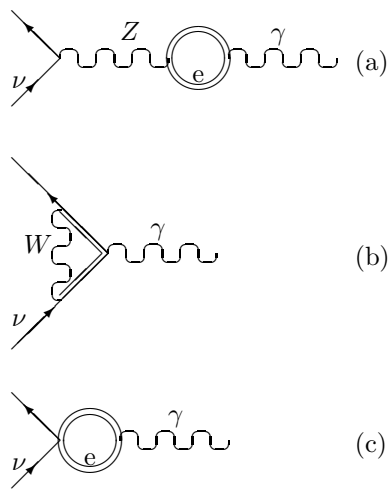
\begin{figure}
\centering\leavevmode
\vbox{
\unitlength=0.8mm
\begin{picture}(60,25)
\put(8,15){\line(-1,1){8}}
\put(8,15){\line(-1,-1){8}}
\put(0,7){\vector(1,1){4}}
\put(8,15){\vector(-1,1){6}}
\multiput(9.5,15)(6,0){3}{\oval(3,3)[t]}
\multiput(12.5,15)(6,0){3}{\oval(3,3)[b]}
\put(31,15){\circle{10}}
\put(31,15){\circle{9}}
\multiput(37.5,15)(6,0){3}{\oval(3,3)[t]}
\multiput(40.5,15)(6,0){3}{\oval(3,3)[b]}
\put(0,10){\shortstack{{}$\nu$}}
\put(18,18){\shortstack{{$Z$}}}
\put(43,18){\shortstack{{$\gamma$}}}
\put(30,11){\shortstack{{e}}}
\put(57,13){\shortstack{{(a)}}}
\end{picture}

\unitlength=0.8mm
\begin{picture}(60,32)
\put(16,15){\line(-1,1){7.5}}
\put(16,15){\line(-1,-1){7.5}}
\put(15,15){\line(-1,1){7}}
\put(15,15){\line(-1,-1){7}}
\put(16,15){\line(-1,1){16}}
\put(16,15){\line(-1,-1){16}}
\put(1,0){\vector(1,1){6}}
\put(16,15){\vector(-1,1){10}}
\multiput(17.5,15)(6,0){3}{\oval(3,3)[t]}
\multiput(20.5,15)(6,0){3}{\oval(3,3)[b]}
\multiput(8.2,8.7)(0,6){3}{\oval(3,3)[l]}
\multiput(8.2,11.7)(0,6){2}{\oval(3,3)[r]}
\put(2,15){\shortstack{{}$W$}}
\put(0,3){\shortstack{{}$\nu$}}
\put(23,18){\shortstack{{$\gamma$}}}
\put(57,13){\shortstack{{(b)}}}
\end{picture}

\vskip3ex

\unitlength=0.8mm
\begin{picture}(60,25)
\put(8,15){\line(-1,1){8}}
\put(8,15){\line(-1,-1){8}}
\put(0,7){\vector(1,1){4}}
\put(8,15){\vector(-1,1){6}}
\put(13,15){\circle{10}}
\put(13,15){\circle{9}}
\multiput(19.5,15)(6,0){3}{\oval(3,3)[t]}
\multiput(22.5,15)(6,0){3}{\oval(3,3)[b]}
\put(0,10){\shortstack{{}$\nu$}}
\put(26,18){\shortstack{{$\gamma$}}}
\put(12,11){\shortstack{{e}}}
\put(57,13){\shortstack{{(c)}}}
\end{picture}
}
\smallskip
\caption[...]{Neutrino-photon coupling in an external magnetic field.
The double line represents the electron propagator in the presence of
a $B$-field. 
(a)~$Z$-$A$-mixing. (b)~Penguin diagram (only for $\nu_e$).
(c)~Effective coupling in the limit of infinite gauge-boson masses.
\label{Fig1}}
\end{figure}

\section{The Neutrino-Photon-Vertex}

In a magnetic field, photons couple to neutrinos by the amplitudes
shown in Figs.~1(a) and (b). The electron propagator, represented by a
double line, is modified by the field to allow for a nonvanishing
coupling. It has been speculated that superstrong magnetic fields may
exist in the early universe, but we limit our discussion to field
strengths not very much larger than $B_{\rm crit}=m_e^2/e$ which is
the range thought to occur in pulsars.  Therefore, while in principle
similar graphs exist for $\mu$ and $\tau$ leptons, we may neglect
their contribution. For the same reason we may ignore field-induced
modifications of the gauge-boson propagators.  Moreover, we are
interested in neutrino energies very much smaller than the $W$- and
$Z$-boson masses, allowing us to use the limit of infinitely heavy
gauge bosons and thus an effective four-fermion interaction,
\begin{equation}
{\cal L}_{\rm eff}= 
-\frac{G_F}{\sqrt{2}}\,\bar{\nu} \gamma_{\mu}(1-\gamma_5)\nu\,
\bar{E}\gamma^{\mu}(g_V-g_A \gamma_5)E.
\end{equation}
Here, $E$ stands for the electron field,
$\gamma_5=i\gamma^0\gamma^1\gamma^2\gamma^3$,  
$g_V=2\sin^2\theta_W+\frac{1}{2}$ and 
$g_A=\frac{1}{2}$ for $\nu_e$, and
$g_V=2\sin^2\theta_W-\frac{1}{2}$ and
$g_A=-\frac{1}{2}$ for $\nu_{\mu,\tau}$.
In our subsequent calculations we will always use
$\sin^2\theta_W=\frac{1}{4}$ for the weak mixing angle so that the
vector coupling will identically vanish for $\nu_\mu$ and 
$\nu_\tau$. Anyhow, we will find that the axial coupling is far more 
important.

The $\nu$-$\gamma$-vertex is then given by the
amplitude shown in Fig.~1(c) for which we find
\begin{eqnarray}
\label{m1}
{\cal M}&=& i\frac{e G_F}{\sqrt{2}}Z\varepsilon_{\mu}
\bar{\nu} \gamma_{\nu}(1-\gamma_5)\nu \\\nonumber
  &\times& \int \frac{d^4 p}{(2\pi)^4}
  {\rm Tr}[\gamma^{\mu}G(p)\gamma^{\nu}
(g_V-g_A \gamma_5)G(p-k)].
\end{eqnarray}
Here, $G(p)$ denotes the electron propagator in a magnetic field, $p$
the four momentum of the electron in the loop, and $k$ the four
momentum of the photon line. Further, $\varepsilon$ is the photon
polarization vector and $Z$ its wave-function renormalization
factor. For the physical circumstances of interest to us, the photon
refractive index will be very close to unity so that we will be able
to use the vacuum approximation $Z=1$.

The matrix element Eq.~(\ref{m1}) can be written in the form
\begin{equation}
\label{m}
{\cal M}=-\frac{G_F}{\sqrt{2}\,e}Z\varepsilon_{\mu}
\bar{\nu}\gamma_{\nu}(1-\gamma_5)\nu\,
(g_V\Pi^{\mu \nu}-g_A\Pi_5^{\mu \nu})
\end{equation} 
where
\begin{eqnarray}
\Pi^{\mu \nu}(k)&=&-ie^2
\int \frac{d^4 p}{(2\pi)^4}
{\rm Tr}[\gamma^{\mu}G(p)\gamma^{\nu} G(p-k)],\\
\Pi_5^{\mu\nu}(k)&=&-ie^2
\int \frac{d^4 p}{(2\pi)^4}
{\rm Tr}[\gamma^{\mu}G(p)\gamma^{\nu}\gamma_5 G(p-k)].
\end{eqnarray} 
Here, $\Pi$ is the photon polarization tensor or vector-vector (VV)
response function in the magnetic field, while $\Pi_5$ is the
vector-axial vector (VA) response function.

In the presence of the external field the electron propagation
function $G(x_1,x_2)$ satisfies the equation
\begin{equation}
\label{dirac}
\left[ m_e + \gamma 
\left( \frac{\partial}{i\partial x_1}-eA(x_1) \right ) \right ] 
G(x_1,x_2) = \delta(x_1-x_2)
\end{equation}
which can be solved exactly by Schwinger's proper-time
method~\cite{Schwinger}.  For the case of a purely homogeneous
magnetic field in the 3-direction ($F_{12}=-F_{21}=B_3=B$) the result
is \cite{Schwinger,Tsai}
\begin{equation}
\label{propagator}
G(x_1,x_2)= \Phi(x_1,x_2) 
\int \frac{d^4p}{(2\pi)^4}e^{ip(x_1-x_2)}G(p).
\end{equation}
Here,
\begin{equation}
\Phi (x_1,x_2)=\exp\left[ie\int_{x_2}^{x_1}dy\,A(y)\right]
\end{equation}
and
\begin{eqnarray}
\label{here}
G(p)&=&i\int_0^{\infty}ds\, 
\exp \left [ -is \left(m_e^2+p_{\|}^2+
\frac{\tan z}{z}\,p_{\bot}^2\right)\right]\nonumber\\ 
&&\hskip3em\times\,\frac{1}{\cos z}\,
\biggl [ (m_e-\gamma p_{\|})e^{i\sigma_3z}-
\frac{\gamma p_{\bot}}{\cos z}\biggr],
\end{eqnarray}
where $z=eBs$. Note that
$\sigma_3z=F_{\mu\nu}\sigma^{\mu \nu}z/2B$ with
$\sigma^{\mu \nu} \equiv \frac{i}{2}[\gamma^{\mu},\gamma^{\nu}]$.
The $\|$ and $\bot$ decomposition of a four vector $a$ is defined by
$a_\|=(a_0,0,0,a_3)$ with a spatial part parallel to the external
$B$-field and $a_\bot=a-a_\|=(0,a_1,a_2,0)$. 
In the absence of a magnetic field ($B\to0$) obviously
$G(p)=(\gamma p +m_e-i0)^{-1}$.

The photon polarization tensor implied by this result has been
calculated in Refs.~\cite{Adler,Tsai,Minguzzi}. Following
Ref.~\cite{Tsai} it is
\begin{eqnarray}
\Pi^{\mu\nu}(k)&=&\frac{e^3B}{(4\pi)^2}
\int_0^{\infty}ds\int_{-1}^{+1} dv\nonumber\\
&\times&
\Bigl\{e^{-is\phi_0}\Bigl[(g^{\mu \nu}k^2-k^{\mu}k^{\nu})N_0
\nonumber\\
&&-\,(g^{\mu \nu}_{\|}k^2_{\|}-k_{\|}^{\mu}k^{\nu}_{\|})N_{\|}+
(g^{\mu\nu}_{\bot}k^2_{\bot}-k^{\mu}_{\bot}k^{\nu}_{\bot})N_{\bot}
\Bigr]\nonumber\\
&&-\,e^{-ism_e^2}(1-v^2)(g^{\mu \nu}k^2-k^{\mu}k^{\nu})\Bigr\},
\end{eqnarray}
where 
\begin{equation}
\phi_0=m_e^2+\frac{1-v^2}{4}\,k_{\|}^2+
\frac{\cos zv -\cos z}{2z \sin z}\,k_{\bot}^2.
\end{equation}
Further, 
\begin{eqnarray}
N_0&=&\frac{\cos zv-v\cot z\,\sin zv}{\sin z},\nonumber\\
N_\|&=&-\cot z\left(1-v^2+\frac{v\sin zv}{\sin z}\right)+
\frac{\cos zv}{\sin z},\nonumber\\
N_\bot&=&-\frac{\cos zv}{\sin z}+\frac{v\cot z\sin zv}{\sin z}
+2\frac{\cos zv-\cos z}{\sin^3 z}.
\end{eqnarray}
The $\|$ and $\bot$ decomposition of the metric is
$g_\|={\rm diag}(-,0,0,+)$ and 
$g_\bot=g-g_\|={\rm diag}(0,+,+,0)$.

The VA response function $\Pi_5$ has been calculated in
Ref.~\cite{Raad}. However, their calculation contains several errors
which require a reconsideration of $\Pi_5$. It is~\cite{Raad}
\begin{eqnarray}
\label{p5n}
\Pi_5^{\mu \nu}(k)&=&
i\frac{e^3}{(4\pi)^2}
\int_0^{\infty}ds \int_{-1}^{+1} dv \,e^{-is\phi_0}
\nonumber\\
&\times& 
\Bigl\{\Bigl(2m^2_e+\frac{1-v^2}{2}k_{\|}^2\Bigr)\widetilde{F}^{\mu \nu}
-\,(1-v^2)k_{\|}^{\nu}(\widetilde{F} k)^{\mu}\nonumber\\
&&\hskip1em+\,
R\Bigl[k_{\bot}^{\mu}(k\tilde{F})^{\nu}
-k_{\bot}^2\widetilde{F}^{\mu \nu}\Bigr]\Bigr\},
\end{eqnarray}
where
\begin{equation}\label{Rdef}
R=\frac{1-v \sin zv \sin z - \cos z \cos zv}{\sin ^2 z}
\end{equation}
and $\widetilde{F}^{\mu \nu}=
\frac{1}{2}\epsilon^{\mu \nu \rho \sigma}F_{\rho \sigma}$
with $\epsilon^{0123}=1$ is the dual of the field-strength tensor.

This result is not gauge invariant. However, one may integrate the
first term under the integral by parts~\cite{Raad}
\begin{eqnarray}
\label{ip}
&&\int_0^\infty ds\,\Bigl(2m^2_e+\frac{1-v^2}{2}k_{\|}^2\Bigr)
e^{-is\phi_0}
\nonumber\\
&&\hskip5em=\,-2i-\int_0^\infty ds
\,k_{\bot}^2 e^{-is\phi_0}R.
\end{eqnarray}
The first term on the r.h.s.\ does not depend on the mass of the
particle in the loop.  For the $Z$-$A$-mixing amplitude (Fig.~1a) it
cancels when we take into account all fermions from each generation
according to the cancellation of the Adler anomaly in the Standard
Model~\cite{Adler69}. For the penguin diagram this term disappears  
when we take into account the exact $W$ propagator~\cite{Wyler}.

With these results we find for the VA response function
\begin{eqnarray}
\label{p5}
\Pi_5^{\mu \nu}(k)&=&\frac{e^3}{(4\pi)^2m_e^2}
\Bigl\{-C_\|\,k_{\|}^{\nu}(\widetilde{F} k)^{\mu}\nonumber\\
&+&C_\bot\,\Bigl[k_{\bot}^{\nu}(k\widetilde{F})^{\mu}
+k_{\bot}^{\mu}(k\widetilde{F})^{\nu}-
k_{\bot}^2\widetilde{F}^{\mu \nu}\Bigr]\Bigr\},
\end{eqnarray} 
where 
\begin{eqnarray}
C_\|&=&im_e^2\int_0^{\infty}ds\int_{-1}^{+1}
dv\,e^{-is\phi_0}(1-v^2)
\nonumber\\
C_\bot&=&im_e^2\int_0^{\infty}ds\int_{-1}^{+1}
dv\,e^{-is\phi_0}R
\end{eqnarray}
are dimensionless coefficients which are real for $\omega<2m_e$, i.e.\
below the pair-production threshold. 

%%%%%%%%%%%%%%%%%%%%%%%%%%%%%%%%%%%%%%%%%%%%%%%%%%%%%%%%%%%%%%%%%%%%%%
%% Section IV %%%%%%%%%%%%%%%%%%%%%%%%%%%%%%%%%%%%%%%%%%%%%%%%%%%%%%%%
%%%%%%%%%%%%%%%%%%%%%%%%%%%%%%%%%%%%%%%%%%%%%%%%%%%%%%%%%%%%%%%%%%%%%%

\section{Cherenkov Rate}

Armed with these results we may now turn to an evalulation of the rate
for $\nu\to\nu\gamma$. It is easy to see that for both photon
eigenmodes the parity-conserving part of the effective vertex
($\Pi^{\mu \nu}$) is proportional to the small parameter
$(n_{\|,\bot}-1)^2 \approx
(\alpha/2\pi)\eta_{\|,\bot}\sin^2\beta$.  It is important to
note that the parity-violating part ($\Pi_5^{\mu \nu}$) is {\it not\/}
proportional to this small parameter for the $\parallel$ photon mode,
while it is proportional to it for the $\perp$ mode.

It is interesting to compare this finding with the standard 
plasma decay process $\gamma\to\bar\nu\nu$ which is dominated by the
VV vertex function. Therefore, in the approximation
$\sin^2\theta_W=\frac{1}{4}$ only the electron
flavor contributes to plasmon decay. Here, we are in the opposite
situation where the axial coupling to the electrons is the dominating
one so that the Cherenkov rate is equal for 
(anti)neutrinos of all flavors.

For neutrinos which propagate perpendicular to the magnetic
field, Eqs.~(\ref{gc}), (\ref{m}), and (\ref{p5}) lead to a 
Cherenkov emission rate of $\parallel$ photons of 
\begin{equation}
\label{cher}
\Gamma=\frac{2\alpha G_F^2}{(4\pi)^4}
\left(\frac{B}{B_{\rm crit}}\right)^2
\int_0^{\omega_{\rm max}}\!\!d\omega\,
\omega^4\left(1-\frac{\omega}{E}\right)
\Bigl(\frac{C_\|}{2}-C_\perp\Bigr)^2\!\!.
\end{equation}

We consider at first neutrino energies below the pair-production
threshold $E<2m_e$. For $\omega<2m_e$ the photon refractive
index~\cite{Adler,Erber} always obeys the Cherenkov condition $n>1$ so
that $\omega_{\rm max}=E$.  Further, it turns out that in the range
$0<\omega< 2m_e$ the expression $C_\|/2-C_\perp$ depends only
weakly on $\omega$ so that it is well approximated by its value at
$\omega=0$.  Therefore, Eq.~(\ref{cher}) can be written in the form
\begin{eqnarray}\label{finalresult}
\Gamma&\approx&\frac{4\alpha G_F^2E^5}{135(4\pi)^4}\,
\left(\frac{B}{B_{\rm crit}}\right)^2 h(B)\nonumber\\
&=&2.0\times10^{-9}~{\rm s}^{-1}~\left(\frac{E}{2m_e}\right)^5
\left(\frac{B}{B_{\rm crit}}\right)^2 h(B),
\end{eqnarray}
where 
\begin{equation}
h(B)\equiv\frac{9}{16}(C_\|-2C_\bot)_{\omega=0}^2.
\end{equation}
Explicitly, this is found to be
\begin{equation}
h(B)= 
\cases{(4/25)\,(B/B_{\rm crit})^4&for $B\ll B_{\rm crit}$,\cr
1&for $B\gg B_{\rm crit}$.\cr}
\end{equation}
Evidently, even if the field strength is around the critical value,
the Cherenkov rate is rather small.

Turning next to the case $E>2m_e$ we note that in the presence of a
magnetic field the electron and positron wavefunctions are Landau
states so that the process $\nu\to\nu e^+e^-$ becomes kinematically
allowed. Therefore, neutrinos with such large energies will
lose energy primarily by pair production rather than by Cherenkov
radiation---for recent calculations see Refs.~\cite{Borisov}.

The Cherenkov effect $\nu\to\nu\gamma$ has been previously calculated
in Refs.~\cite{Galtsov,Skobelev76}. However, they have not taken the
neutral-current part into account so that their result applies only to
$\nu_e$ for which the effective axial coupling $g_A=1$ was used
instead of our $g_A=1-\frac{1}{2}=\frac{1}{2}$ which is a sum of the
charged- and neutral-current contributions. Further, their final
result is larger by a factor $2^4\pi$ relative to our
Eq.~(\ref{finalresult}).

We may also compare our $\nu\to\nu\gamma$ calculation with previous
$\nu\to\nu'\gamma$ ones~\cite{Gvozdev,Skobelev95,Wunner}.  To this end
we note that in $\nu\to\nu'\gamma$ the photon energy obeys
$0<\omega<E$ if $\nu$ is ultrarelativistic, if $m_{\nu'}=0$, and if
one uses the vacuum photon dispersion relation. As this is the same
range allowed in our case, there is no phase-space complication and we
may compare our Eq.~(\ref{finalresult}) with Eq.~(15) of
Ref.~\cite{Wunner}.  However, we must identify $\nu'$ with $\nu$ which
implies that we must drop their mixing-angle factors. Further, we must
substitute $g_A=\frac{1}{2}$ for their value 1 to account for the
neutral-current contribution. After these modifications our result is
still a factor of 2 larger. This is explained, we believe, by their
use of an unpolarized initial state of massive Dirac neutrinos while
our neutrinos are always left-handed. 

If $\nu'$ is not taken to be massless, the $\nu\to\nu'\gamma$ decay
rate is reduced and vanishes for $m_{\nu'}=m_\nu$. This, of course, is
just our case of identical initial and final states where the
Cherenkov rate by no means vanishes.  This discrepancy is due to the
use of the vacuum photon dispersion relation in
Refs.~\cite{Gvozdev,Skobelev95,Wunner}. Using a consistent dispersion
relation prevents this suppression effect because for
ultrarelativistic initial neutrinos, the allowed range of photon
energies will always be $0<\omega<E$.

%%%%%%%%%%%%%%%%%%%%%%%%%%%%%%%%%%%%%%%%%%%%%%%%%%%%%%%%%%%%%%%%%%%%%%
%% Section V %%%%%%%%%%%%%%%%%%%%%%%%%%%%%%%%%%%%%%%%%%%%%%%%%%%%%%%%%
%%%%%%%%%%%%%%%%%%%%%%%%%%%%%%%%%%%%%%%%%%%%%%%%%%%%%%%%%%%%%%%%%%%%%%

\section{Summary and Conclusions}

We have calculated the neutrino Cherenkov process
\hbox{$\nu\to\nu\gamma$} in a homogeneous magnetic field. The magnetic
field provides an effective $\nu$-$\gamma$-vertex, and it modifies the
photon dispersion relation such that the Cherenkov condition is met
for photon energies $\omega<2m_e$. The neutrino emits primarily
photons with a polarization vector parallel to the transverse
component of the magnetic field (the $\parallel$ propagation
eigenmode), and the coupling is primarily due to the VA (pseudotensor)
electromagnetic vertex function. We have corrected some errors of a
previous calculation of this dominant term which had been studied in
the context of the $\gamma\to\bar\nu\nu$ process in magnetic fields.
We have also corrected errors in previous calculations of the
Cherenkov process.

For neutrinos propagating transverse to the magnetic field, the
Cherenkov rate is numerically given in Eq.~(\ref{finalresult}). The
strongest magnetic fields known in nature are near pulsars. However,
they have a spatial extent of only tens of kilometers. Therefore, even
if the field strength is as large as the critical one, most neutrinos
escaping from the pulsar or passing through its magnetosphere will not
emit Cherenkov photons. Thus, the magnetosphere of a pulsar is quite
transparent to neutrinos as one might have expected.

%%%%%%%%%%%%%%%%%%%%%%%%%%%%%%%%%%%%%%%%%%%%%%%%%%%%%%%%%%%%%%%%%%%%%%
%% Acknowledgments %%%%%%%%%%%%%%%%%%%%%%%%%%%%%%%%%%%%%%%%%%%%%%%%%%%
%%%%%%%%%%%%%%%%%%%%%%%%%%%%%%%%%%%%%%%%%%%%%%%%%%%%%%%%%%%%%%%%%%%%%%

\section*{Acknowledgments}

We are grateful to Gagik Grigorian, Jose Valle, and Daniel Wyler for
useful discussions. We thank Alexander Kuznetsov and Nicolay Mikheev
for pointing out a missing factor of 2 in Eq.~(24) of our original
manuscript.  A.I.\ acknowledges the hospitality of the
Max-Planck-Institut f\"ur Physik during a visit when this work was
begun. This research was supported, in part, by the Deutsche
Forschungsgemeinschaft grant SFB 375 (G.R.) and by the Lady Davis
Trust (A.I.).

%%%%%%%%%%%%%%%%%%%%%%%%%%%%%%%%%%%%%%%%%%%%%%%%%%%%%%%%%%%%%%%%%%%%%%
%% References %%%%%%%%%%%%%%%%%%%%%%%%%%%%%%%%%%%%%%%%%%%%%%%%%%%%%%%%
%%%%%%%%%%%%%%%%%%%%%%%%%%%%%%%%%%%%%%%%%%%%%%%%%%%%%%%%%%%%%%%%%%%%%%

\end{document}